\documentclass{emulateapj}

\newcommand{\Om}{\mbox{$\Omega_{\mbox{\scriptsize $M$}}$}}
\newcommand{\Ol}{\mbox{$\Omega_{\mbox{\scriptsize $\Lambda$}}$}}

\newcommand{\chandra}{{\it Chandra}}

\newcommand{\rtfh}{\mbox{$r_{\mbox{\scriptsize 2500}}$}}
\newcommand{\rfh}{\mbox{$r_{\mbox{\scriptsize 500}}$}}
\newcommand{\MR}{\mbox{$M_{\mbox{\scriptsize GL}}$}}
\newcommand{\MRs}{\mbox{$M_{\mbox{\scriptsize GL,sph}}$}}
\newcommand{\MRH}{\mbox{$M_{\mbox{\scriptsize HSE}}$}}
\newcommand{\YR}{\mbox{$Y$}}
\newcommand{\MY}{\MR$-$\YR}
\newcommand{\Msun}{M_\odot}

\defcitealias{SmithE05}{S05}

\shorttitle{LoCuSS: SZE and Lensing Measurements of Clusters}
\shortauthors{Marrone et al.}
\slugcomment{The Astrophysical Journal, 701:L114-L118, 2009
August 20}

\begin{document}

\title{LoCuSS: A Comparison of Sunyaev-Zel'dovich Effect and
  Gravitational Lensing Measurements of Galaxy Clusters}

\author{
  Daniel P.\ Marrone,$\!$\altaffilmark{1,2,3}
  Graham P.\ Smith,$\!$\altaffilmark{4}
  Johan Richard,$\!$\altaffilmark{5,6}
  Marshall Joy,$\!$\altaffilmark{7}
  Massimiliano Bonamente,$\!$\altaffilmark{8}
  Nicole Hasler,$\!$\altaffilmark{8}
  Victoria Hamilton-Morris,$\!$\altaffilmark{4}
  Jean-Paul Kneib,$\!$\altaffilmark{9}
  Thomas~Culverhouse,$\!$\altaffilmark{2,3} 
  John E.\ Carlstrom,$\!$\altaffilmark{2,3,10,11}
  Christopher~Greer,$\!$\altaffilmark{2,3} 
  David~Hawkins,$\!$\altaffilmark{12}
  Ryan~Hennessy,$\!$\altaffilmark{2,3} 
  James~W.~Lamb,$\!$\altaffilmark{12}
  Erik~M.~Leitch,$\!$\altaffilmark{2,3} 
  Michael~Loh,$\!$\altaffilmark{2,9}
  Amber~Miller,$\!$\altaffilmark{13,14}
  Tony~Mroczkowski,$\!$\altaffilmark{13,15}
  Stephen~Muchovej,$\!$\altaffilmark{12,15}
  Clem~Pryke,$\!$\altaffilmark{2,3,11}
  Matthew~K.~Sharp,$\!$\altaffilmark{2,10} 
  David~Woody\altaffilmark{12}
}

\altaffiltext{1}{Jansky Fellow, National Radio Astronomy Observatory}
\altaffiltext{2}{Kavli Institute for Cosmological Physics, Department
  of Astronomy and Astrophysics, University of Chicago, Chicago, IL 60637, USA}
\altaffiltext{3}{Department of Astronomy and Astrophysics, University
  of Chicago, Chicago, IL 60637, USA} 
\altaffiltext{4}{School of Physics and Astronomy, University of
  Birmingham, Edgbaston, Birmingham, B15 2TT, UK} 
\altaffiltext{5}{California Institute of Technology, Mail Code 105-24,
  Pasadena, CA 91125, USA} 
\altaffiltext{6}{Department of Physics, Durham University, South Road,
  Durham, DH1 3LE, UK} 
\altaffiltext{7}{Space Science Office, VP62, NASA/Marshall Space
  Flight Center, Huntsville, AL 35812, USA}  
\altaffiltext{8}{Department of Physics, University of Alabama,
  Huntsville, AL 35812, USA} 
\altaffiltext{9}{Laboratoire d'Astrophysique de Marseilles, OAMP,
  CNRS-Universit\'e Aix-Marseilles, 38 rue Fr\'ed\'eric Joliot-Curie,
  13388 Marseilles Cedex 13, France} 
\altaffiltext{10}{Department of Physics, University of Chicago,
  Chicago, IL 60637, USA} 
\altaffiltext{11}{Enrico Fermi Institute, University of Chicago,
  Chicago, IL 60637, USA} 
\altaffiltext{12}{Owens Valley Radio Observatory, California Institute
  of Technology, Big Pine, CA 93513, USA}  
\altaffiltext{13}{Columbia Astrophysics Laboratory, Columbia
  University, New York, NY 10027, USA} 
\altaffiltext{14}{Department of Physics, Columbia University, New York,
  NY 10027, USA} 
\altaffiltext{15}{Department of Astronomy, Columbia University, New
  York, NY 10027, USA} 

\begin{abstract}
   We present the first measurement of the relationship between the
   Sunyaev-Zel'dovich effect (SZE) signal and the mass of galaxy
   clusters that uses gravitational lensing to measure cluster mass,
   based on 14 X-ray luminous clusters at $z\simeq0.2$ from the Local
   Cluster Substructure Survey.  We measure the integrated Compton
   $y$-parameter, $Y$, and total projected mass of the clusters (\MR)
   within a projected clustercentric radius of 350~kpc, corresponding
   to mean overdensities of 4000$-$8000 relative to the critical
   density.  We find self-similar scaling between \MR\ and $Y$, with a
   scatter in mass at fixed $Y$ of 32\%.  This scatter exceeds that
   predicted from numerical cluster simulations, however, it is
   smaller than comparable measurements of the scatter in mass at
   fixed $T_X$.  We also find no evidence of segregation in \YR\
   between disturbed and undisturbed clusters, as had been seen with
   $T_X$ on the same physical scales.  We compare our scaling relation
   to the \citeauthor{BonamenteE08} relation based on mass
   measurements that assume hydrostatic equilibrium, finding no
   evidence for a hydrostatic mass bias in cluster cores (\MR\ =
   $0.98\pm0.13$ \MRH), consistent with both predictions from
   numerical simulations and lensing/X-ray-based measurements of
   mass-observable scaling relations at larger radii.  Overall our
   results suggest that the SZE may be less sensitive than X-ray
   observations to the details of cluster physics in cluster cores.
\end{abstract}


\setcounter{footnote}{15}
\section{Introduction}\label{s:intro}

The Sunyaev-Zel'dovich effect (SZE) is a weak distortion of the cosmic
microwave background (CMB) spectrum introduced as CMB photons
propagate through foreground galaxy clusters \citep{SZ70,SZ72}.
Because this signal does not suffer cosmological dimming and is
expected to closely track cluster mass, the SZE is a potentially
powerful tool for producing large, mass-limited galaxy cluster samples
that can be used to constrain dark energy, under the proviso that the
relationship between cluster mass and the observed signal is well
calibrated \citep{CHR02}.  

The relationship between the SZE observable, the integrated Compton
$y$-parameter, $Y$, and cluster mass is difficult to calibrate because
of the difficulty of measuring mass directly. To date all examinations
of the mass-$Y$ relationship have assumed hydrostatic equilibrium
(HSE) when converting SZE or X-ray measurements of the intracluster
gas into estimates of the total mass \citep[\MRH;
e.g.][]{MorandiE07,BonamenteE08}. These methods measure scaling
relations between closely related quantities derived from the same
data; deviations in one parameter will therefore inevitably be
correlated with the other, and the true intrinsic scatter will likely
be underestimated. Furthermore, both simulations
\citep[e.g.][]{NagaiE07,JeltemaE08,PiffarettiValdarnini08,LauE09} and
observations \citep{MahdaviE08} suggest that the HSE assumption is
incorrect; non-thermal sources of pressure support may bias \MRH\ or
increase the scatter with $Y$.

The scatter in mass at fixed $Y$ is an important quantity to determine
in preparation for the SZE cluster surveys designed to constrain the
dark energy equation of state, as scatter in the mass-observable
relation is generally degenerate with the cosmological parameters of
interest \citep[e.g.][]{SmithE03,StanekE06}. These surveys will rely
on self-calibration techniques to build cluster mass functions from
SZE-selected cluster catalogs, however these techniques can be
rendered ineffective if the scatter is much larger than the few
percent predicted by simulations
\citep{daSilvaE04,MotlE05,Nagai06}. At present there are only weak
observational limits, but independent measurements of the mass-$Y$
scatter can be used to refine the self-calibration
\citep{LimaHu05}.

The Local Cluster Substructure Survey
(LoCuSS\footnote{http://www.sr.bham.ac.uk/locuss}; G. P. Smith et
al. 2009, in preparation) is a morphologically unbiased survey of
X-ray luminous galaxy clusters in a narrow redshift range
($0.15<z<0.3$). The survey unites a variety of probes of total mass,
intracluster gas, and cluster member galaxies (e.g.,
\citealt{SmithE05}, hereafter \citetalias{SmithE05};
\citealt{ZhangE08}; \citealt{OkabeE09}). One of the main goals of the
survey is to measure the shape, normalization and scatter of cluster
mass-observable relations as an input to cosmological cluster studies.
Here we use a pilot sample of clusters, assembled from
\citetalias{SmithE05}, \citet{BonamenteE06}, and early LoCuSS
observations with the Sunyaev-Zel'dovich Array (SZA), to compare the
SZE signal with cluster masses derived from gravitational lensing
(\MR) and thus to construct the first mass-$Y$ relation that is
independent of the assumption of HSE. We summarize
the observations and modeling in Section~\ref{s:obs}. The \MY\
relationship, scatter, and its implications are discussed in
Section~\ref{s:res}. We assume \Om=0.3, \Ol=0.7, and $h$=0.7; the median
cluster redshift is $z=0.222$, at which 1\arcsec\ corresponds to a
physical scale of 3.6~kpc.

\section{Observations and Modeling}
\label{s:obs}

\begin{deluxetable*}{lccccccc}
\tablecolumns{8}
\tablewidth{0pt}
\tablecaption{Cluster Sample\label{t:cls}}
\tablehead{Cluster & Redshift & $Y$ & $YD_\mathrm{A}^2$ & SZE & 
 $M_\mathrm{GL}$ & Lensing & Classification\\ & &
 $(10^{-10})$ & $(10^{-5} \mathrm{Mpc}^2)$ & Ref.\ & $(10^{14} M_\odot)$ & Ref. & }
\startdata
A\,68   & 0.255 & 0.55$\pm$0.08 & 3.67$\pm$0.52 & 1 & 3.48$\pm$0.07 & 2,3 & Disturbed \\
A\,209  & 0.206 & 0.94$\pm$0.14 & 4.57$\pm$0.67 & 4 & 1.23$\pm$0.39 & 2 & Disturbed \\
A\,267  & 0.230 & 0.53$\pm$0.06 & 3.08$\pm$0.34 & 1 & 2.20$\pm$0.34 & 2 & Disturbed \\
A\,383  & 0.188 & 0.39$\pm$0.04 & 1.61$\pm$0.16 & 4 & 3.71$\pm$0.82 & 2 & Undisturbed \\
A\,611  & 0.288 & 0.39$\pm$0.04 & 3.13$\pm$0.34 & 1 & 2.12$\pm$0.05 & 5 & Undisturbed \\
A\,773  & 0.217 & 1.03$\pm$0.11 & 5.40$\pm$0.57 & 1 & 4.03$\pm$0.12 & 2,5 & Disturbed \\
Z\,2701 & 0.214 & 0.28$\pm$0.03 & 1.46$\pm$0.16 & 4 & 1.92$\pm$0.07 & 5 & Disturbed \\
A\,1413 & 0.143 & 1.83$\pm$0.26 & 4.90$\pm$0.70 & 1 & 2.59$\pm$0.50 & 5 & Undisturbed \\
A\,1689 & 0.181 & 1.86$\pm$0.15 & 7.51$\pm$0.60 & 1 & 7.44$\pm$0.05 & 6 & Disturbed \\
A\,1763 & 0.288 & 0.56$\pm$0.06 & 3.10$\pm$0.32 & 4 & 1.42$\pm$0.54 & 2 & Disturbed \\
A\,1835 & 0.253 & 1.03$\pm$0.07 & 6.82$\pm$0.48 & 1 & 3.35$\pm$0.06 & 2,5 & Undisturbed \\
A\,2218 & 0.171 & 1.12$\pm$0.10 & 4.23$\pm$0.38 & 1 & 4.23$\pm$0.09 & 2 & Disturbed \\
A\,2219 & 0.228 & 1.12$\pm$0.05 & 6.27$\pm$0.26 & 4 & 3.48$\pm$0.07 & 2 & Disturbed \\
A\,2537 & 0.297 & 0.42$\pm$0.03 & 3.47$\pm$0.24 & 4 & 1.75$\pm$0.90 & 7 & Disturbed \\
\enddata
\tablerefs{(1) \citet{BonamenteE08}; (2) \citetalias{SmithE05}; (3)
  \citet{RichardE07}; (4) this work; (5) Richard et al. 2009, in
  preparation; (6) \citet{LimousinE07}; (7) Hamilton-Morris et al. 2009, in
  preparation. } 
\tablecomments{\YR\ and \MR\ measured at 350~kpc radius. Morphological
classification from lensing and X-ray data.  See \citetalias{SmithE05}
and \citet{SmithE08}.}
\end{deluxetable*}

We analyze SZE and gravitational lensing observations of 14 clusters,
listed in Table~\ref{t:cls}.  SZE results for eight of these come from
\citet{LaRoqueE06} and \citet{BonamenteE06,BonamenteE08}, the
observational details being found in \citeauthor{LaRoqueE06}. New SZE
measurements of A\,209, A\,383, A\,1763, A\,2219, A\,2537, and Z\,2701
were obtained at 31~GHz using the SZA, which has baselines of
350$-$1300$\lambda$ for SZE sensitivity (angular scales of
110\arcsec-7\arcmin) and 2-7.5~k$\lambda$ for radio source removal. A
detailed description of SZA observations and analysis methods can be
found in
\citet{MuchovejE07}. Typical integration times were $\sim30-40$~hr
per cluster, and the rms noise per beam $\sim0.2$mJy. In six of 14
clusters we detect radio sources within an arcminute of the cluster
center, these sources were presented in \citet{CobleE07}.

Our interferometric observations do not measure directly the total SZE
flux of the cluster within an aperture, this must be derived from a
model cluster profile.  As in previous works, we have modeled these
clusters with an isothermal $\beta$-model, which has been shown to
provide $Y$ measurements indistinguishable from those from a
simulation-motivated pressure profile for radii smaller than
$r_{vir}/3$ \citep{MroczkowskiE09}.  The shape parameters for the
$\beta$-model and an isothermal X-ray temperature were jointly derived
from the SZE data and from \chandra\ X-ray images after applying a
100~kpc core cut \citep{LaRoqueE06}.  Best-fit cluster parameters were
obtained using the Markov chain Monte Carlo method described in
\citet{BonamenteE04}. The fluxes of detected radio sources and
central sources present in the 1.4~GHz NVSS catalog
\citep{CondonE98} but undetected at 31~GHz are included as parameters
in the Markov chains and marginalized over.

Details of the gravitational lensing mass measurements are described
by \citetalias{SmithE05} and \citet{RichardE07}; the source of each
measurement is listed in Table~\ref{t:cls}.  Briefly, the mass
measurements are based on \emph{Hubble Space Telescope} (\emph{HST})
imaging of the cluster cores, ground-based spectroscopy of
multiply imaged galaxies identified in these data, and by
weakly sheared background galaxies.  These data constrain
parameterized models of the projected mass distribution in the cluster
cores comprising one or more cluster-scale mass components (to
describe the dark matter and intracluster gas) plus typically $\sim30$
galaxy-scale mass components per cluster. Measurements of the projected
cluster mass and the uncertainty on the mass are then obtained by
integrating the mass distributions of the family of models within the
chosen confidence interval in parameter space (in this letter we quote
errors at 68\% confidence) out to the chosen radius.

\citetalias{SmithE05} \citep[see also][]{SmithE08} used the structure of
the cluster mass distributions inferred from these lens models -- the
substructure fraction, $f_{\rm sub}$ -- in conjunction with
\emph{Chandra} observations to classify the cluster cores as
``disturbed'' or ``undisturbed''. The most straightforward criterion
in this classification was the offset between the peaks of the X-ray
emission and the lensing-based mass map, in the sense that a
statistically significant offset indicates that the cluster core is
disturbed in some way, probably due to a cluster-cluster merger
\citep[see also][]{SandersonE09}.  We adopt the classification of
\citet{SmithE08} in this Letter and apply it to the five additional
clusters in our sample (Table~\ref{t:cls}).

\section{Results and Discussion}
\label{s:res}

\subsection{The Mass-$Y$ Scaling Relation}

\begin{figure}
\plotone{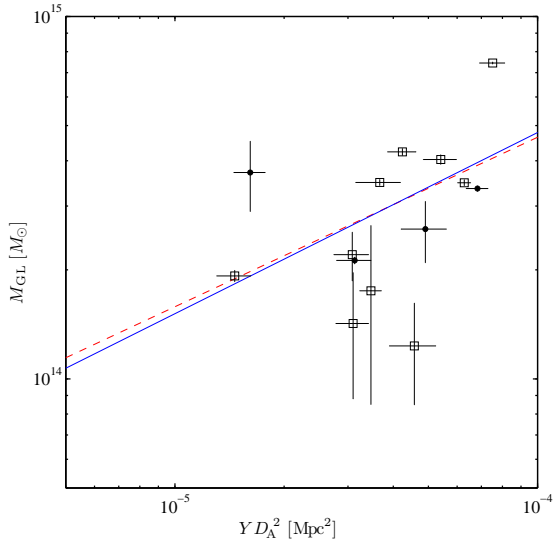}
\caption{Scaling between projected mass (\MR) based on gravitational
lens models and \YR, both measured within a projected clustercentric
radius of 350~kpc.  Clusters classified as undisturbed in
Table~\ref{t:cls} are shown as filled circles, disturbed clusters are
shown as open boxes. The solid line shows the scaling relation when
the slope is fixed to the self-similar value ($\beta=0.5$), the dashed
line is obtained when the slope is also a free parameter.  }
\label{f:m_sz}
\end{figure}

We first define the aperture within which \MR\ and \YR\ are measured.
After some experimentation we chose to retain the aperture used by
\citetalias{SmithE05}, a clustercentric radius of 250$h^{-1}$~kpc, or
350~kpc in our adopted cosmology. This fixed physical aperture is
well-matched to the \emph{HST}/WFPC2 and Advanced Camera for Surveys
(ACS) fields of view, and is somewhat larger than the typical
resolution of our SZE data. The shape of our $\beta$-model fit is
jointly determined from both SZA and high-resolution X-ray data and
thus is well resolved. \MR\ and \YR\ values measured at this radius
are listed in Table~\ref{t:cls} and plotted in Figure~\ref{f:m_sz}.

We fit our projected mass (\MR) and \YR\ data points in the log(\YR
$D_\mathrm{A}^2$)-log(\MR) plane, where a power law scaling relation
takes the form $\mathrm{log}\left(M_{GL}\right) =
\alpha+\beta~\mathrm{log}\left(YD_\mathrm{A}^2\right)$.  Starting with
the self-similar scaling \citep{Kaiser86}, it can be shown that the
relation between mass and $Y$ within a fixed physical radius has a
slope $\beta=0.5$ with no dependence on the redshift evolution of the
Hubble parameter.  We seek to measure the slope and normalization of
the scaling relation, as well as the intrinsic scatter between these
parameters, and therefore follow the prescription of \citet{WeinerE06}
for linear regression in the presence of intrinsic scatter. This
Bayesian method includes the intrinsic scatter in the $y$-coordinate
($\sigma_y$) as a parameter of the fit and maximizes the probability
of the model ($\alpha$, $\beta$, $\sigma_y$) given the data and
errors. The likelihood takes the form (Equation A3 of
\citealt{WeinerE06})
\begin{equation}
\mathcal{L}=-\Sigma\frac{\left[y_i-\left(\alpha+\beta x_i\right)\right]^2}{\beta^2e_{xi}^2+e_{yi}^2+\sigma_y^2},
\end{equation}
where the $e_i$ are the $x$ or $y$ errors in the individual data
points.

We perform the regression in two different ways: first, with both the
normalization ($\alpha$) and slope ($\beta$) of the fit free we obtain
$\alpha=16.5^{+1.1}_{-0.9}$ and $\beta=0.47^{+0.24}_{-0.20}$, with an
intrinsic scatter in mass at fixed $Y$ of $\sigma_{M|Y}=32\pm4$\%.
With the slope fixed to the self-similar value ($\beta=0.5$) we obtain
$\alpha=16.68\pm0.04$, again with $\sigma_{M|Y}=32\pm4$\%.  Despite
large uncertainties, the \MY\ slope is consistent with
self-similarity.  For comparison, \citet{BonamenteE08} and
\citet{MorandiE07} found that the \MRH$-$\YR\ scaling relation is
consistent with self-similarity at \rtfh\ (the radius where the
average interior density is $\Delta$=2500 times the critical density
of the universe, typically 30$-$100\% larger than the radius used
here).  In contrast, \citetalias{SmithE05} were unable to constrain
the slope of the \MR$-T_X$ relation, suggesting that the intrinsic
scatter in \MR-$Y$ is smaller than in \MR-$T_X$.

\subsection{Intrinsic Scatter}\label{s:intsc}

We now compare the intrinsic scatter in the \MY\ relation with that in
\MR$-$$T_X$ from \citetalias{SmithE05}.  We first use our regression method
to re-fit their data, taking the opportunity to adopt the theoretical
slope ($T_X=BM_\mathrm{GL}^{1/\beta}$, $\beta=1$, in their
nomenclature) appropriate for this scaling relation within a fixed
aperture. With their 10 clusters we find a scatter in mass of
$\sigma_{M|T}=41\pm5$\%, which is 28\% larger (1.5$\sigma$
significance) than $\sigma_{M|Y}$. Indeed, both $\sigma_{M|Y}$ and
$\sigma_{M|T}$ are large compared to the scatter predicted from
numerical simulations \citep[e.g.][]{Nagai06}, likely due to a
combination of modeling uncertainties, astrophysical processes in
cluster cores, and projection effects. The effects of modeling
uncertainties are hard to quantify but should not be ignored, as both
\MR\ and \YR\ are derived from parameterized models. To the extent
that these models inadequately represent the cluster or underestimate
the uncertainty in the derived parameters, our inferred scatter will
be artificially increased. Thus our measured $\sigma_{M|Y}$ is an
upper limit to the true scatter between mass and $Y$ in cluster cores.

In general, cluster mass measurements based on gravitational lensing
and SZE observations are more susceptible to projection effects than
those based on X-ray data because the density-squared dependence of
the X-ray emissivity limits contributions from material along the line
of sight.  Nevertheless, simulations \citep{HolderE07,ShawE08} suggest
that the contribution to the SZE signal from projection is unimportant
for clusters with $M_{vir}>10^{14}h^{-1}\Msun$, with the projection
effects diminishing further at the small radii used
here. Gravitational lensing is sensitive to all mass along the line of
sight through the cluster, which will likely increase the scatter in
our scaling relation, however the contribution of correlated
large-scale structure to lensing-based mass measurements should
decrease at smaller radii in a manner similar to the impact on the SZE
signal.  We also emphasize that our sample is X-ray selected with no
consideration paid to the presence/absence of strong lensing arcs, so
it should not suffer the halo orientation bias that boosts the core
mass in strong-lensing-selected samples
\citep[e.g.][]{HennawiE07}. Indeed, 5/14 clusters contain no obvious
strong-lensing signal. Projection-induced scatter therefore appears
insignificant for \YR\ and may be somewhat more important for \MR,
although this affects both $\sigma_{M|Y}$ and $\sigma_{M|T}$.

Given that the mass measurement methods and 9/14 of the clusters are
identical between this Letter and \citetalias{SmithE05}, we expect the
difference between the scatter in these two relations to be dominated
by the relative sensitivity of $Y$ and $T_X$ to the gas physics of
cluster cores.  The intracluster medium (ICM) in cluster cores is
often disturbed by active galactic nucleus (AGN) activity
\citep[e.g.][]{FabianE06}, and cluster-cluster mergers
\citep{MarkevitchVikhlinin07}. For this reason, cluster cores are
often excluded when deriving X-ray observables for use as mass proxies
\citep[e.g.][]{ArnaudE05,KravtsovE06,Maughan07}. We therefore ascribe
the difference between $\sigma_{M|Y}$ and $\sigma_{M|T}$ to the
difference in the SZE and X-ray emissivity dependencies on ICM density
and temperature.  The SZE is unaffected by isobaric disturbances
(excluding relativistic components such as cosmic rays), while the
energy-integrated X-ray emissivity varies approximately as the square
of the density, which may be significantly perturbed in the
core. \citet{PfrommerE07} found potentially important changes in the
SZE and X-ray signals from cool cores in simulations incorporating
cosmic rays, but the effect and therefore the induced scatter was
found to be larger for X-ray emission. In larger measurement
apertures, such as the frequently used \rtfh\ and \rfh\ (350~kpc
corresponds to $\Delta$=4000$-$8000 for our sample), core-removed
X-ray mass proxies are found to correlate more tightly with mass
\citep[e.g.][]{VikhlininE09-CCCP2}. We expect that $\sigma_{M|Y}$ will
also decrease when measured on these scales.

\subsection{Structural Segregation}

\citetalias{SmithE05} found that the normalization of the \MR$-T_X$
relation for disturbed clusters was $\sim40\%$ hotter than for
undisturbed clusters at $\sim2.5\sigma$ (after excising the central
region of the X-ray data).  We re-fit the \MR-\YR\ relation to the
undisturbed and disturbed sub-samples (Table~\ref{t:cls}) finding no
significant difference in normalization between undisturbed and
disturbed clusters
($\alpha_{undisturbed}-\alpha_{disturbed}=0.0\pm0.1$, $\beta$ fixed to
0.5).  Again, this suggests that the integrated SZE signal, even in
cluster cores, is less sensitive to the ICM physics than the X-ray
temperature, although the significance of this result is low given the
small samples used here and in \citetalias{SmithE05}.

\subsection{Hydrostatic Equilibrium}

As discussed in Sections \ref{s:intro} and \ref{s:intsc}, the ICM may
be far from HSE; we search for this effect in our data by combining
our results with those of \citet{BonamenteE08} who studied the
\MRH-\YR\ relation.  Our lensing measurements are sensitive to the
projected mass ($M_{cyl}$); we therefore convert them to the spherical
masses ($M_{sph}$) required for comparison with the hydrostatic
analysis.  We do this by assuming that the halo mass distributions are
described by the Navarro-Frenk-White (NFW) profile \citep{NFW}, which
is specified by mass and concentration, $M_{vir}$ and $c_{vir}$; given
\MR\ and $c_{vir}$ we can than derive an approximate conversion factor
$M_{cyl}/M_{sph}$ at 350~kpc. This conversion factor depends
sensitively on $c_{vir}$, but there is considerable theoretical
uncertainty about the mean value and scatter in concentration at fixed
mass, and its variation with redshift and cluster mass. Without direct
measurements of $c_{vir}$ for most of our clusters we marginalize over
the range of values observed by \citet{OkabeE09} for 19 clusters from
the LoCuSS sample, which have very similar masses and redshifts to our
sample. We randomly choose a concentration from the 19 in
\citet{OkabeE09} for each of our clusters, calculate the corresponding
value of \MRs=\MR$\times M_{sph}/M_{cyl}$, and fit a scaling relation
between the estimated \MRs\ and \YR.  We repeat this process thousands
of times, and derive a mean normalization of the \MRs$-$$Y$ relation
of $\alpha_{\mathrm{GL},sph}=16.42\pm0.05$ for $\beta=0.5$.  This
value can be directly compared to the normalization of the \MRH$-$$Y$
relation from \citet{BonamenteE08}:
$\alpha_\mathrm{HSE}=16.425\pm0.016$.  The statistically insignificant
difference between these values implies a mass ratio of
\MRs/\MRH=$0.98\pm0.13$.

Using hydrostatic masses from X-ray observations and weak lensing
masses from the literature, \citet{ZhangE08} and
\citet{VikhlininE09-CCCP2} estimate the weak lensing to hydrostatic
mass ratio to be $1.09\pm0.08$ and $1.01\pm0.11$, respectively,
measured at \rfh. \citet{MahdaviE08}, however, demonstrate a radial
trend in this ratio between \rtfh\ and \rfh, suggestive of a 20\%
deficit in hydrostatic masses at \rfh\ that cannot be accounted for by
the projection of unrelated structures along the line of sight.  Our
result, interior to the innermost radius examined by
\citet{MahdaviE08}, is consistent with their non-detection of a
bias in the hydrostatic mass at small radii.  The precision of our
measurement is limited by the intrinsic scatter in our relation, with
a smaller contribution coming from the unknown halo concentration
values required to convert between \MR\ and \MRs.  Future combinations
of weak and strong lensing measurements, specifically joint
strong$+$weak lens modeling of the clusters combining the
\emph{Subaru} and \emph{HST} data (G. P. Smith et al. 2009, in
preparation) should allow us to evaluate the bias as a function of
radius.

\section{Summary}

We have presented the first calibration of the mass-$Y$ relation based
on gravitational-lensing measurements of cluster mass, based largely
on previously published SZE data and gravitational lens models.  We
construct the relation at a fixed physical radius of 350kpc to
minimize uncertainties arising from extrapolation of cluster models
based on both datasets.  In contrast to the mass-$T_X$ relation within
the same aperture \citepalias{SmithE05}, we succeed in fitting a model
with both free slope and normalization.  The best-fit slope is
consistent with the self-similar prediction, and the intrinsic scatter
in mass at fixed $Y$ is 32\%, in contrast to the 41\% scatter in mass
at fixed $T_X$.  We also fit the relation to sub-samples of disturbed
and undisturbed cluster cores but find that the best-fit
normalizations for these sub-samples are consistent within the errors.
Finally, we combine our results with those of \citet{BonamenteE08} to
test whether the cluster cores are in HSE. Cluster core masses
estimated from lensing and SZE data (assuming HSE) are, on average,
consistent within the errors, suggesting that departures from
equilibrium are modest in cluster cores.  We conclude that the
difference between the mass-$Y$ and mass-$T_X$ relations is mainly
attributable to the relative insensitivity of the SZ effect to the
physics of the ICM in cluster cores. Future articles will explore the
mass-$Y$ relation at larger radii through weak lensing and in larger
samples.

\section*{Acknowledgments}

We thank our colleagues in the LoCuSS collaborations for much support,
encouragement and help.  GPS acknowledges support from the Royal
Society and STFC, and thanks the Kavli Insitute of Cosmological
Physics at the University of Chicago for their hospitality whilst
working on this Letter.  GPS thanks Alain Blanchard for helpful
comments. We gratefully acknowledge the James S.\ McDonnell
Foundation, the National Science Foundation and the University of
Chicago for funding to construct the SZA. The operation of the SZA is
supported by NSF Division of Astronomical Sciences through grant
AST-0604982. Partial support is provided by NSF Physics Frontier
Center grant PHY-0114422 to the Kavli Institute of Cosmological
Physics at the University of Chicago, and by NSF grants AST-0507545
and AST-05-07161 to Columbia University.

{\it Facilities:} \facility{SZA}, \facility{HST (WFPC2)}

\end{document}